\documentclass[conference]{IEEEtran}
\IEEEoverridecommandlockouts
\usepackage{cite}
\usepackage{amsmath,amssymb,amsfonts,latexsym}
\usepackage{amsthm}
\usepackage{braket}
\usepackage{graphicx}
\usepackage{textcomp}
\usepackage{xcolor}
\usepackage{mathtools}
\usepackage{multirow}
\usepackage{url}
\usepackage{booktabs}
\usepackage{hyperref}
\usepackage{lipsum}
\usepackage{nicematrix}
\usepackage[all]{xy}
\usepackage{tikz-cd}
\usepackage{tikz}
\usepackage{tikz-3dplot}
\usepackage{makecell}
\usepackage{caption}
\usepackage{subcaption}
\usepackage{quantikz}
\usepackage{soul}

\usetikzlibrary{calc,hobby}
\usetikzlibrary{intersections,shapes.arrows}
\usetikzlibrary{arrows.meta, positioning, backgrounds}

\usepgfmodule{nonlineartransformations}
\def\fluttertransform{%
    \pgfgetlastxy\x\y
    \pgfpoint{\x+sin(\y)}{\y+sin(\x)*(30-\x/2)+\x/10}
}

\definecolor{mfold}{RGB}{96,86,196}

\def \vphi{\varphi}

\def \a{\alpha}

\def \mcal{\mathcal}

\newtheorem{theorem}{Theorem}
\newtheorem*{theorem*}{Theorem}
\newtheorem*{lemma*}{Lemma}
\newtheorem*{example*}{Example}
\newtheorem{corollary}{Corollary}
\newtheorem*{remark*}{Remark}

\newtheorem{observation}{Observation}

\newenvironment{sketchproof}
  {\begin{proof}[Sketch of proof]}
  {\end{proof}}

\newcommand{\C}{\mathbb{C}}
\newcommand{\U}{\mathcal{U}}
\newcommand{\End}{\operatorname{End}}
\newcommand{\GL}{\operatorname{GL}}
\newcommand{\Ad}{\operatorname{Ad}}
\newcommand{\Ker}{\operatorname{Ker}}

\newcommand{\dagg}{^{\dagger}}

\DeclareMathOperator{\dimR}{dim_{\mathbb{R}}}
\DeclareMathOperator{\dimC}{dim_{\mathbb{C}}}

\makeatletter
\renewenvironment{proof}[1][\proofname]{%
  \par\pushQED{\qed}\normalfont%
  \topsep6\p@\@plus6\p@\relax%
  \trivlist\item[\hskip\labelsep\bfseries #1\@addpunct{.}]\ignorespaces
}{%
  \popQED\endtrivlist\@endpefalse
}
\makeatother

\def\BibTeX{{\rm B\kern-.05em{\sc i\kern-.025em b}\kern-.08em
    T\kern-.1667em\lower.7ex\hbox{E}\kern-.125emX}}

\makeatletter
\newcommand{\linebreakand}{%
  \end{@IEEEauthorhalign}
  \hfill\mbox{}\par
  \mbox{}\hfill\begin{@IEEEauthorhalign}
}
\makeatother

\title{Observable Geometry for Effective Quantum Circuits\thanks{
The views expressed in this article are those of the authors and do not represent the views of Wells Fargo. This article is for informational purposes only. Nothing contained in this article should be construed as investment advice. Wells Fargo makes no express or implied warranties and expressly disclaims all legal, tax, and accounting implications related to this article.}}

\author{Huan-Hsin Tseng$^1$, Hsin-Yi Lin$^1$, Samuel Yen-Chi Chen$^2$, Yan Mong Chan$^3$, Tzu-Chieh Wei$^3$, Shinjae Yoo$^1$\\
\small $^1$AI \& ML Department, Brookhaven National Laboratory, Upton NY, USA\\
\small $^2$Wells Fargo, New York NY, USA\\
\small $^3$Department of Physics \& Astronomy, C. N. Yang Institute for Theoretical Physics, Stony Brook University, Stony Brook NY, USA\\
\small $^1$\texttt{\{htseng, hlin7, sjyoo\}@bnl.gov}\\
\small $^2$\texttt{ycchen1989@ieee.org}\\
\small $^3$\texttt{\{yanmong.chan, tzu-chieh.wei\}@stonybrook.edu}
}

\begin{document}
\maketitle

\begin{abstract}
We study redundancy and effectiveness of Variational Quantum Circuits via algebraic and geometric views of Lie groups. Considering unitary transformations acting on Hermitian observables, a stabilizer group decomposition is given. Subsequently, we identify the Hermitian orbit with a quotient space of a symmetric space. Through this connection, we characterize the effective circuit degrees of freedom. Our approach of spectral decompositions and homogeneous spaces yields tractable calculation criteria, which are verified by numerical experiments.
\end{abstract}

\section{Introduction}

Variational Quantum Algorithms (VQAs)~\cite{cerezo2021variational} are a principal approach for optimizations using near-term quantum processors. Examples include the Variational Quantum Eigensolver (VQE)~\cite{peruzzo2014variational}, Quantum Approximate Optimization Algorithms~\cite{farhi2014quantum}, and Variational Quantum Circuits (VQC)~\cite{chen2020VQDQN}.

Their practical performance depends on various factors, such as hardware and ansatz choices. Although increasing the number of gates may enlarge circuit expressivity, it may also introduce noise, redundant parameters, and trainability pathologies such as barren plateaus~\cite{mcclean2018barren, haug2021capacity}. %

From an engineering perspective, redundant gates inflate depth, sampling cost, and optimizer overhead. A cause of the redundancy in circuits is due to the degeneracy of a Hamiltonian giving constant energy, which results in gauge freedom. Viewing from a mathematical perspective, internal symmetries of a Hamiltonian generate invariant subspaces to stabilize the quantum expectation value as outputs.

Adaptive and architecture-search methods address the trade-off by selecting more economical ansatz structures \cite{grimsley2019adaptive, du2022quantum, kuo2021quantum}. Nevertheless, a systematic understanding of these gauge symmetries remains desirable, as it provides structural criteria for removing observable-invisible unitary directions from variational ansatz design.

This study analyzes the internal symmetries induced by the observable and the corresponding stable structure and homogeneous spaces towards an effective circuit space.

\section{Problem Setup}

Typical VQAs, including VQEs, VQCs, QAOAs, solve tasks by seeking unitaries extremizing a given Hamiltonian $H$ via $U \mapsto U^{\dagger} H U$. However, in this setting, not all circuits effectively explore the Hamiltonian \textit{orbit}. 

For example, two different unitaries $U_1 = c \, U$, $U_2 = \frac{1}{c} \, U$ up to a constant $c \in \C$, $| c |^2 = 1$ of another unitary $U$ may result in the same effect $U_1^{\dagger} H U_1 = U_2^{\dagger} H U_2$. This tells us certain degrees of freedom in a unitary matrix are redundant in view of $H$, and they ought to be removed from a circuit set for efficiency and effectiveness in quantum optimization. 

\begin{figure}[htbp]
    \centering 
\begin{tikzpicture}[scale=0.7]

\begin{scope}[yshift=20mm]
\pgftransformnonlinear{\fluttertransform}
\draw [fill=green!15] plot [smooth cycle]
coordinates {(-1.14,-1)(-0.84, -.18) (-0.04, 0.3) (2.24, 0)
(4.48, -0.56) (4.48, -1.46) (3.38,-1.84)(0.38, -1.28)};
\end{scope}

    \coordinate (p1) at (0.1, 1.5);
    \coordinate (p2) at (2.3, 1.8);

    \filldraw (4.5, 2.3) node[left] {$\mathbb{H}(n)$};
    
    \filldraw (p1) circle (1.5pt) node[below] {\footnotesize{$H_0$}};
    \filldraw (p2) circle (1.5pt) node[below] {\footnotesize{$H_1$}};
    
    \draw[->, dashed] (p1) to[bend left=25] node[midway, above] {\footnotesize{$U$}} (p2);

\begin{scope}[xshift=84mm, yshift=11mm]
  \begin{scope}[rotate=-6]
    \shadedraw[draw=mfold!75!black, line width=1.4pt, line join=round,
               inner color=mfold!8, outer color=mfold!38, closed hobby]
      plot coordinates {
        ( 2.68, 0.22)   %
        ( 1.40, 1.06)   %
        (-0.70, 1.30)   %
        (-1.78, 0.82)   %
        (-1.42, 0.20)   %
        (-2.1,-0.58)   %
        ( 0.05,-1.22)   %
        ( 1.78,-0.74)   %
      };
  \end{scope}

\node[mfold!80!black] at (1.8, 1.3) {$\mathcal{U}(V)$};

\coordinate (U1) at (-0.80,-0.15);   %
\coordinate (U2) at ( 1.60, 0.32);   %

\draw[black, line width=0.9pt]
  (U1) .. controls (0.20,0.38) and (1.00,0.56) .. node[midway, above left] {$S$} (U2);

\filldraw[black] (U1) circle (1.5pt) node[below] {$U_1$};
\filldraw[black] (U2) circle (1.5pt) node[below] {$U_2$};
\end{scope}

\end{tikzpicture}
\caption{[Left] Motions in observable space $\mathbb{H}(n)$ via a unitary. [Right] A subset of $\U(V)$, giving a constant in map $U \mapsto U\dagg H U$ for fixed $H$ as (\ref{E: const map}), is identified by $S$ in (\ref{E: stabilizer group}).}
\label{fig: Unitary transformations}
\end{figure}
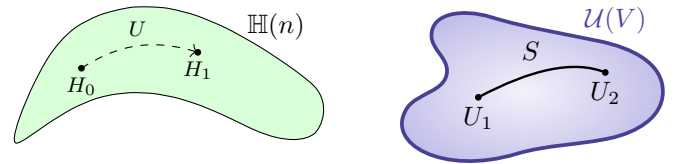

This redundancy is a form of gauge freedom, where distinct unitaries correspond to the same observable effect. We formalize our inquiry:

\textbf{Question:}
\textit{For two Hermitians $H_0 \neq H_1$, is there a unique unitary $U$ such that $U\dagg H_0 U = H_1$   (\textbf{Fig.~\ref{fig: Unitary transformations}} [Left])? If not, how far is the choice of $U$ from being unique? (\textbf{Fig.~\ref{fig: Unitary transformations}} [Right])}

To probe the uniqueness, assume unitaries $U_1, U_2$ such that
\begin{equation}\label{E: const map}
      U_1\dagg H U_1 = U_2\dagg H U_2
\end{equation}
Denoting $A:=U_1 U_2\dagg $, we see $  A\dagg H A=H$, or equivalently, $[H,A] = 0$ as $A$ is also unitary. This observation leads us to the following consideration. 

Let $(V, \langle \, \cdot \, \rangle )$ be a finite-dimensional Hilbert space and $\U(V):= \{ U \in \GL(V) \, | \, \langle U v, U w \rangle = \langle v, w \rangle, \, \forall v, w \in V \}$ is the unitary group of $V$. For $m$-qubit system, $V = (\C^2)^{\otimes m}$.

Given a Hermitian $H$, define 
\begin{equation}\label{E: stabilizer group}
    S = \{ A \in \U(V) \, | \, [H, A] = 0\}
\end{equation}
and consider the quotient space $S \backslash \U(V) := \{ S u \, | \, u \in \U(V) \}$, where $S u$ (also denoted as $[u]$) is a \textit{right coset} of $S$ and $u \in \U(V)$. We reserve $\U(V) / S$ for the \textit{left coset} space. Note that $S$ defined by (\ref{E: stabilizer group}) depends on $H$ and forms a subgroup of $\U(V)$, but may \textit{not} be normal ($U S U^{-1} \subseteq S$ for all $U \in \U(V)$), so that $S \backslash \U(V)$ does not have group structures in general. Then,

\begin{theorem}\label{Thm: disjoint Hermitian orbits}
For distinct cosets $[u_1] \neq [u_2]$ in $S \backslash \U(V)$,
\[
    [u_1\dagg] \, H [u_1] \, \neq [u_2\dagg] \, H \, [u_2]
\]
\end{theorem}

\begin{sketchproof}
Suppose otherwise. Let $[u_1\dagg] H [u_1] = [u_2\dagg] H [u_2]$; so $u_2 \, u_1\dagg \, S\dagg H S \,  u_1 \, u_2 \dagg = S\dagg H S $. Because $S$ commutes with $H$, we know $S\dagg H S = H$, or elementwisely $A\dagg H A = H$ for all $A \in S$. Then, $u_2 u_1\dagg H  u_1 u_2 \dagg = H$, forcing $u_1 u_2 \dagg \in S$ or $u_1 \in S u_2$. But this  contradicts to $[u_1] \neq [u_2]$.
\end{sketchproof}

Therefore, Theorem~\ref{Thm: disjoint Hermitian orbits} precisely answers our question above, where the elements in $S \backslash \U(V)$ uniquely push one Hermitian to another, and $S$ is the redundant component to be removed from $\U(V)$.

While the answer of $S \backslash \U(V)$ is theoretically elegant, computing $S$ from (\ref{E: stabilizer group}) is not easy since all unitaries commuting with $H$ are to be collected. For this, we need a tractable form of $S$ by analyzing the composition.

\section{Structural Decompositions}

Let spectral decomposition of Hermitian $H \in \End_\C V$ be
\begin{equation}\label{E: H decomp}
      H = \lambda_1 T_1 + \lambda_2 T_2 + \cdots + \lambda_k T_k,
\end{equation}
where $\lambda_1, \ldots, \lambda_k \in \C$ are the distinct eigenvalues of $H$, and $T_j: V \to W_j$ is an orthogonal projection onto eigenspace $W_j:=\Ker(H - \lambda_j I) \subseteq W_j$ for each $j=1, \ldots, k$ such that
\[
  V=W_1\oplus W_2\oplus\cdots\oplus W_k, \quad \sum_{j=1}^k T_j=I, \quad T_iT_j=\delta_{ij}T_j.
\]
Then $S$ has the following decomposition,
\begin{theorem}\label{Thm: Stabilizer decomp}
    \begin{equation}\label{E: Stabilizer decomp}
    \begin{aligned}
        S &= \U(W_1)\times\U(W_2)\times\cdots\times\U(W_k), \quad \text{with}\\
        \U(W_j) &:= \{ A: W_j \to W_j \, | \, \langle A w_1, A w_2 \rangle =  \langle w_1, w_2 \rangle \}
    \end{aligned}
    \end{equation}
    the unitary group of eigenspace $W_j \subseteq V$ of $\lambda_j$.
\end{theorem}

\begin{sketchproof}
Let $A \in \U(W_1) \times \cdots \times \U(W_k)$ and $A = A_1 + \cdots + A_k$ with $A_j\in\U(W_j)$.
Since $V = \bigoplus_j W_j$, write $v = w_1 + \cdots + w_k$ for any $v \in V$ with $w_j\in W_j$, and
\[
\begin{aligned}
  HAv &= H \left( A_1 w_1 + \cdots + A_kw_k \right) = \lambda_1 A_1 w_1 + \cdots + \lambda_k A_k w_k,\\
  AHv &= A(\lambda_1 w_1 + \cdots + \lambda_k w_k) = \lambda_1 A_1 w_1 + \cdots + \lambda_k A_k w_k.
\end{aligned}
\]
Hence $H A = A H$, so $A \in S$. Conversely, let $A \in S$. With decomposition (\ref{E: H decomp}) we have,
\[
  \left(\sum_{\ell = 1}^k \lambda_\ell T_\ell \right) A = A \left( \sum_{\ell = 1}^k \lambda_\ell T_\ell \right).
\]
Acting $T_i$ on the left and $T_j$ on the right yields,
\[
( \lambda_i - \lambda_j ) \, T_i A \, T_j = 0
\]
Whence $T_i A \, T_j = 0$ for distinct eigenvalues $i \neq j$. Therefore
\begin{equation}\label{E: AT commute}
      AT_i= \left(\sum_{\ell = 1}^k T_\ell \right) A T_j = T_j A \, T_j = T_\ell A \left(\sum_{\ell=1}^k T_\ell \right) = T_j A.
\end{equation}
where $\sum_\ell T_\ell = I$ is used. Thus, $A(W_j) \subseteq W_j$ and so it makes sense to define the restriction $A_j := A |_{W_j} : W_j \to W_j$. To show that $A_j$ is unitary, notice $A_j = T_j \circ A$ and compute
\[
\langle A_j w, A_j w' \rangle = \langle (T_j \circ A) w, (T_j \circ A) w' \rangle
\]
for any $w, w' \in W_j$. Then by (\ref{E: AT commute}) we see,
\[
\langle ( A \circ T_j) w,  ( A \circ T_j) w' \rangle = \langle A w,  A w' \rangle =  \langle w, w' \rangle.
\]
Thus $A = A_1 \oplus \cdots \oplus A_k$ and $A_j \in \U(W_j)$.
\end{sketchproof}

Theorem~\ref{Thm: Stabilizer decomp} then provides us with a tractable form for calculating $S$ instead of (\ref{E: stabilizer group}) as each eigenspace is computable. Next, we show this result can be used to estimate the ``size'' of the effective Hamiltonian by all possible circuits. For this, some established tools from \textit{homogeneous spaces} will be used.

\section{Characterizing effective Hamiltonian via homogeneous spaces}

Recall the Lie group action. Let $G$ be a Lie group and $M$ be a smooth manifold. Denote $M$ a \textbf{$G$-space} if it is equipped with a (smooth) function (also called an \textit{action}) $\vphi: G \times M \to M$ by $(g, p) \mapsto \vphi(g, p)$ with $\vphi(e, p) = p$ for all $p \in M$ and identity $e \in G$. We consider the \textit{right} action in Quantum physics, $g \mapsto \vphi_g := \vphi(g, \, \cdot \, )$ such that $\vphi_{g_1 g_2} = \vphi_{g_2} \circ \vphi_{g_1}$ for $g_1, g_2 \in G$. Note that the order of $g_1, g_2$ is reversed in compositions.

The \textit{orbit} of $p \in M$ by $(G, \vphi)$ is defined as the set $ \vphi(G, p) := \{ \vphi(g, p) \in M \, | \, \text{for all $g \in G$} \}$. The \textbf{stabilizer group} (also called \textit{isotropy group} or \textit{little group}) of $p \in M$ collects the elements fixing $p$ by $G_p := \{ g \in G \, | \, \vphi(g, p) = p \}$.

A smooth manifold $M$ is further called a \textbf{homogeneous} $G$-space if the action is also \textit{transitive}, that is, for any $p, q \in M$, there exists $g \in G$ such that $\vphi(g, p) = q$.

Under the group action language, $S \subseteq \U(V)$ in (\ref{E: stabilizer group}) can be viewed as a \textbf{stabilizer}. In our setting, let $M = \End_{\C}(V)$ and $G = \U(V)$. Defining $\vphi: G \times M \to M$ by $g \mapsto \vphi_g := \Ad_{g^{-1}}$ gives a right action with $\Ad_g(H) := g H g^{-1} = g H g\dagg$ the adjoint operator. Indeed, for $g_1, g_2 \in \U(V)$, $H \in \End_{\C}(V)$,
\[
    \Ad_{(g_1 g_2)^{-1}}(H) = (g_1 g_2)^{-1} H (g_1 g_2) =  \left( \Ad_{g_2^{-1}} \circ \Ad_{g_1^{-1}} \right) (H)
\]
Therefore, we see the stabilizer of a Hermitian $H$ under $g \mapsto \Ad_{g^{-1}}$ is exactly (\ref{E: stabilizer group}),
\begin{equation}
    G_H = \{ U \in \U(V) \, | \, \Ad_{U^{-1}}(H) = H \} = S
\end{equation}

With this view of group actions, the effects of quantum circuits on observable $H$ can be interpreted with geometric intuition by some results from Lie groups as follows.

Recall that if $B \subseteq G$ is a Lie subgroup, then one may define an equivalence relation on $G$ by
\[
  g_1 \sim g_2 \,\, \Leftrightarrow \,\, g_1 g_2^{-1} \in B \,\, \Leftrightarrow \,\, g_1 = A g_2 \,\, \text{for some } A \in B,
\]
to form a quotient space collecting the right cosets of $B$ in $G$,
\begin{equation}\label{E: Lie quotient group}
    B\backslash G := \{ R_g B \, | \, g \in G \} =\{ R_g A \, | \, g \in G, \, A \in B\}
\end{equation}
where $R_g h := h g$ is the usual group multiplication of $R: G \times G \to G$ with $g \mapsto R_g := R(g, \cdot)$. As such, $R$ can be viewed as a right action onto $G$ itself, or $B \backslash G$ in particular. Note that $\vphi_g$ and $R_g$ are two \textit{distinct} right actions of $G$ on manifold $M$ and group $G$, respectively.

The following theorem gives a homogeneous space construction of Lie groups.
\begin{theorem}[Homogeneous space construction~\cite{Lee2012}]\label{Thm: Lie group homogeneous space}
If $B\subseteq G$ is a closed Lie subgroup, then $B\backslash G$ in (\ref{E: Lie quotient group}) is a smooth manifold of dimension $\dim(B \backslash G) = \dim G - \dim B$ admitting a unique smooth structure such that the quotient map $\pi: G \to B \backslash G$ by $\pi(g) = R_g B$ is a smooth submersion. The right multiplication $R: G \times (B \backslash G) \to B \backslash G$ of $G$ acting smoothly on $B \backslash G$ by
\[
  (R_{g_2} \circ R_{g_1} ) B = R_{g_1 g_2} B, \, \, g_1, g_2 \in G
\]
is transitive. Hence $B \backslash G$ is a homogeneous $G$-space.
\end{theorem}

In our setting, one may show the stabilizer $S \subseteq \U(V)$ defined by (\ref{E: stabilizer group}) is a \textit{closed} Lie subgroup. Thus,
\begin{corollary}
    $S \backslash \U(V)$ is a homogeneous $G$-space.
\end{corollary}

\begin{remark*}
    $S \backslash \U(V)$ in this setting is close to the stabilizer construction in Quantum Error Corrections, where $\mathcal{N}(S) \backslash \U(V)$ becomes the syndrome set after replacing $S$ with its normalizer $\mathcal{N}(S)$ and when $\mathcal{N}(S) \vartriangleleft \U(V)$.
\end{remark*}

The purpose of this study is to investigate the space of effective Hamiltonian $U \mapsto U\dagg H U$ given all possible circuits $U$, which is a common ground for VQAs (VQEs, VQCs, etc).

For this goal, consider the orbit of a Hermitian $H \in \End_\C(V)$ under action $\vphi_g = \Ad_{g^{-1}}$, $g \in \U(V)$,
\begin{equation}\label{E: Hermitian orbit}
    M_H = \{ \vphi_U (H) = U\dagg H U \, | \, \forall U \in \U(V) \}
\end{equation}
One sees that $M_H$ is a homogeneous $\U(V)$-space as $\vphi_U$ is transitive. Then the following theorem allows us to convert the study of orbit space $M_H$ into that of Lie groups.

\begin{theorem}[Homogeneous space characterization~\cite{Lee2012}]\label{Thm: Homogeneous characterization}
Let $G$ be a Lie group and let $M$ be a homogeneous $G$-space with right action $\vphi: G \times M \to M$. For any $p \in M$. The stabilizer group $G_p$ is a closed subgroup of $G$, and the map $F : G_p \backslash G \to M$ defined by
\begin{equation}\label{E: equivariant diff}
    F(R_gG_p) = \vphi_g (p)
\end{equation}
is an equivariant diffeomorphism.
\end{theorem}

Since $M_H$ is a homogeneous $G$-space, Theorem~\ref{Thm: Homogeneous characterization} tells us,

\begin{corollary}\label{Cor: equivariant diffeomorphic}
   $S \backslash \U(V) \overset{F}{\cong} M_H$ with equivariant diffeomorphism $F$ defined by (\ref{E: equivariant diff}).
\end{corollary}

Finally, Theorem~\ref{Thm: Stabilizer decomp}, \ref{Thm: Lie group homogeneous space} and Corollary~\ref{Cor: equivariant diffeomorphic} together indicates the effective Hamiltonian space~(\ref{E: Hermitian orbit}) is a manifold of dimension,
\begin{equation}\label{E: orbit dim}
      \dimR M_H = \dimR \U(V) - \dimR S = n^2 - \left( \sum_{j=1}^k n_j^2 \right)
\end{equation}
where $n = \dimC V$, $n_j = \dimC W_j$ with $\sum_{j=1}^k n_j=n$.

(\ref{E: orbit dim}) explicitly depicts that the versatility of effective Hamiltonian space (orbit) depends on given $H$ via the size of the stabilizer $S$ in (\ref{E: stabilizer group}). Two useful observations can be made.

\begin{observation}[Total degeneracy] 
If $H =\lambda I$, $\lambda \in \mathbb{R}$, then $k=1$, $n_1 = n$ by (\ref{E: H decomp}), and $\dimR M_H = n^2 - n^2 = 0$.

In this case, $M_H=\{U \dagg H U\, | \, U \in \U(V) \}=\{\lambda I\}$ is indeed a single point of dim 0 regardless of the circuits applied.
\end{observation}

\begin{observation}
Since Theorem~\ref{Thm: Stabilizer decomp} requires  $\sum_{j=1}^k n_j = n$, by Cauchy-Schwarz we know $\sum_{j=1}^k n_j^2 \geq  \frac{n^2}{k}$
to yield an upper bound for the orbit manifold dimension,
\begin{equation}\label{E: orbitfold dim upper bound}
    \dimR M_H \leq n^2 - n
\end{equation}
The equality of Cauchy-Schwarz occurs when $n_j \equiv 1$ for all $j$, which is exactly the case when $H$ is non-degenerate,
\[
  H = \operatorname{diag}(\lambda_1, \ldots, \lambda_n), \quad \lambda_i \neq \lambda_j \text{ for $i\neq j$}.
\]
\end{observation}

The above investigation shows that the structure of a Hamiltonian determines the effective geometric degrees of freedom. Thus, the task observable should be analyzed beforehand whenever possible to avoid redundant circuit optimization.

\section{Experiments}\label{Sec: Exps}

We first develop a numerical score characterizing the overlap between a gate set $\mcal{G}$ and the $H$-stabilizer.

\subsection{Stabilizer-overlap score}

Let $H \in \End_\C (V)$ be the task Hamiltonian with decomposition (\ref{E: H decomp}) and $S$ be the stabilizer in (\ref{E: stabilizer group}). Denote the Lie algebras of $\U(V)$ and $S$ as $\mathfrak{u}(V) := \{A \in \End_{\C}(V) \, | \, A\dagg + A = 0 \}$, $\mathfrak{s} := \{ A \in\mathfrak{u}(V) \, | \, [H, A] = 0 \}$. Since circuit generators are usually in Hermitian forms, we define the \textit{Hermitian stabilizer algebra} $i \mathfrak{s} := \{ i A \, | \, A \in \mathfrak{s} \}$. By Theorem~\ref{Thm: Stabilizer decomp}, we may define a projection $ \Pi_{i\mathfrak{s}}: \{A \in \End_{\C}(V) \, | \, A \dagg = A \} \to i\mathfrak{s}$ by,
\begin{equation}\label{eq:stabilizer-algebra-projection}
    \Pi_{i\mathfrak{s}}(A) = \sum_{j=1}^{k} T_j A T_j .
\end{equation}
For Hermitian $A \neq 0$, we define the \textbf{stabilizer-overlap score},
\begin{equation}\label{eq:stabilizer-overlap-score}
    s(A) := \frac{\|\Pi_{i\mathfrak{s}}(A) \|_F^2}{ \| A \|_F^2} \in [0, 1],
\end{equation}
with $\| B \|_F^2 := \operatorname{Tr}(B\dagg B)$. When $s(A) = 1$, $t \mapsto \exp(i t A)$ entirely falls into $S$ to be $[0] \in S \backslash \U(V)$, giving no change to the terminal cost. When $s(A)  = 0$, $[H, \exp(i t A)] \neq 0$ so that it is more visible along $M_H$. %

\subsection{Hamiltonian and selected Lie-algebra candidates}
Consider an $8$-qubit VQE on $V = (\C^2)^{\otimes 8}$ with Hamiltonian
\begin{equation}\label{exp: Hamiltonian}
    H= \sum_{(i,j) \in \mathcal{E}} J_{ij} D_{ij} + \sum_{( (i,j), (k, \ell) ) \in \mathcal{F}} K_{ij, k\ell} D_{ij} D_{k \ell},
\end{equation}
with coefficients $J_{ij}, K_{ij, k\ell} \in \C$, $0 \leq i < j\leq 7$, and \textit{spin-spin coupling} $ D_{ij} := X_i X_j + Y_i Y_j + Z_i Z_j$ with $\{ X_i, Y_i, Z_i \}$ Pauli operators on qubit $i$. This Hamiltonian is non-diagonal in the computational basis and contains four-body Pauli terms. Its degeneracy is protected by global $SU(2)$ symmetry, \textit{i.e.,}
\[
    [H, A_X] = [H, A_Y] = [H, A_Z] = 0
\]
for $A_{\a} := \frac{1}{2} \sum_{j=0}^{7} \a_j$ and $\a \in \{ X, Y, Z \}$. 

We consider 4 families of generators as \textbf{circuit candidates},
\[
 \mcal{G}_1 = \left\{ \a_i  \right\} , \,\, \mcal{G}_2 =  \left\{ \a_i \, \a_j \right\}, \,\, \mcal{G}_3 = \left\{  \sum_{i = 0}^{7} \a_i \right\}, \,\, \mcal{G}_4 =  \left\{ D_{ij} \right\}
\]
for $0 \leq i < j\leq 7$, where they consist of \textit{1-qubit Pauli}, \textit{2-qubit Pauli}, \textit{collective Pauli}, and \textit{spin-spin coupling} generators with family size $|\mcal{G}_1| = 24$, $|\mcal{G}_2|= 84$, $|\mcal{G}_3| = 3$, and $|\mcal{G}_4| = 28$, giving a diverse gate set $\mcal{G} = \bigcup_{i=1}^4 \mcal{G}_i$ of total 139 candidates.

\begin{figure}[htb]
    \vskip -0.1in
    \centering
    \includegraphics[width=1\linewidth]{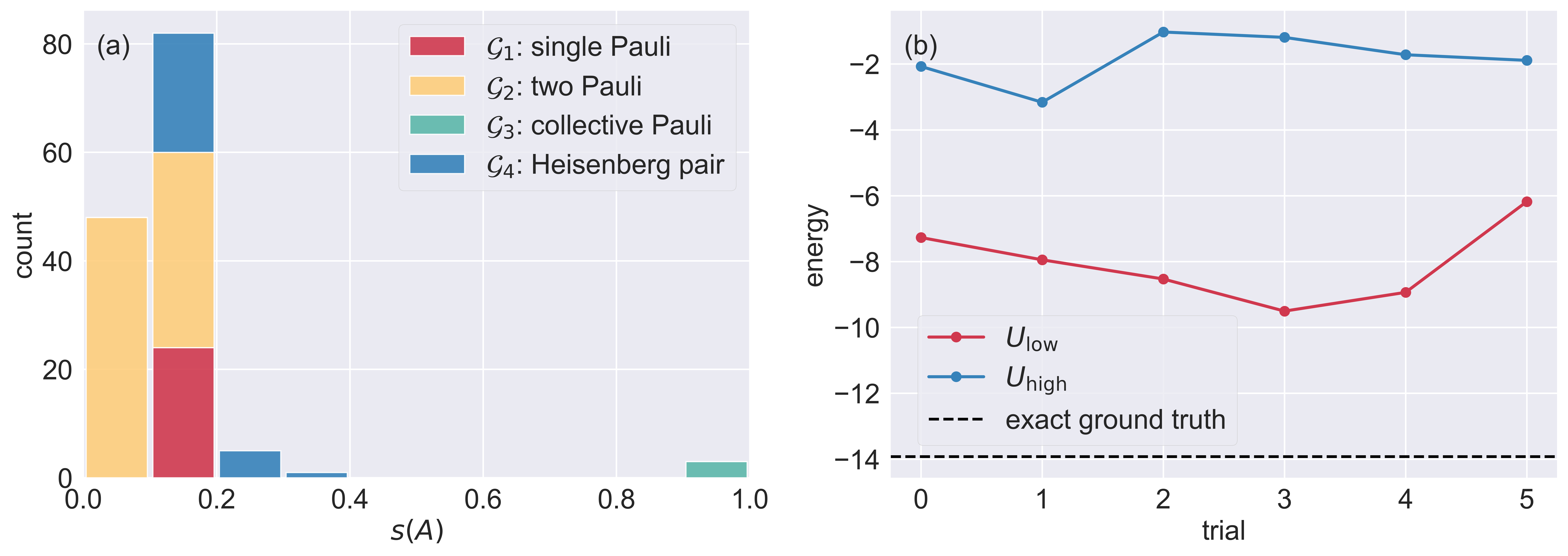}
    \caption{\textbf{(a)} Distribution of $s(A)$ for 4 families $A \in \mcal{G}_i$. \textbf{(b)} VQE energy $E_{\eta}(\boldsymbol{\theta})$ for each trial. $U_{\mathrm{low}}$ reaches lower energy than $U_{\mathrm{high}}$ (dashed line: ground state $E_0 \approx -13.92$).}
    \label{fig: VQE exp}
    \vskip -0.1in
\end{figure}

Each generator $ A \in \mathcal{G}$ determines a one-parameter subgroup of $\U(V)$, or rotations, by $R_A : \mathbb{R} \to \U(V)$, $\theta \mapsto \exp \left( - \frac{i \theta}{2} A \right)$. \textbf{Table}~\ref{tab:pool-family-stats} summarizes some family statistics and \textbf{Fig.}~\ref{fig: VQE exp}(a) shows the distribution of scores $s(\mcal{G})$ for each family pool. Most candidates fall between $0.08 \leq s(A) \leq 0.18$, $A \in \mcal{G}$.

\begin{table}[htb]
    \centering
    \begin{tabular}{lrrrr}
        \toprule
        Circuit family & $| \mcal{G}_i|$ & Avg $s(A)$ & min $s(A)$ & Max $s(A)$ \\
        \midrule
        $\mcal{G}_1$: single Pauli     & 24 & 0.1579 & 0.1423 & 0.1723 \\
        $\mcal{G}_2$: 2-Pauli        & 84 & 0.1041 & 0.0771 & 0.1811 \\
        $\mcal{G}_3$: collective Pauli & 3  & 1.0000 & 1.0000 & 1.0000 \\
        $\mcal{G}_4$: spin-spin  & 28 & 0.1725 & 0.1100 & 0.3355 \\
        \bottomrule
    \end{tabular}
    \caption{Statistics of the circuit pool.}
    \vskip -0.1in
    \label{tab:pool-family-stats}
\end{table}

\subsection{Ans\"atz construction and optimizations}
Reordering elements in $\mathcal{G}$ as $ \{ \overline{A}_1, \overline{A}_2, \ldots, \overline{A}_K \}$ by descending stabilizer-overlap score such that $s( \overline{A}_i) \geq s( \overline{A}_j)$ for $i < j$, we take the \textit{first} $r$ and \textit{last} $r$ elements to form two ans\"atze,
\begin{equation}\label{eq:low-high-ansatz}
\begin{aligned}
     U_{\mathrm{high}}(\boldsymbol{\theta}) &:= R_{\overline{A}_r^{\eta}}(\theta_r) \, \cdots \,  R_{\overline{A}_1^{\eta}}(\theta_1) \\
     U_{\mathrm{low}}(\boldsymbol{\theta}) &:= R_{\overline{A}_K^{\eta}}(\theta_r) \, \cdots \,  R_{\overline{A}_{K - r + 1}^{\eta}}(\theta_1)
\end{aligned}
\end{equation}
where $\boldsymbol{\theta} = (\theta_1, \ldots, \theta_r) \in \mathbb{R}^r$ denotes their variational parameters. $U_{\mathrm{high}}$ and $U_{\mathrm{low}}$ then represent two variational circuits of high and low stabilizer overlap. For $r=24$, $U_{\mathrm{low}}$ consists entirely of 2-qubit Pauli rotations and has mean score $0.0836$ with range $[0.0771, 0.0883]$, while $U_{\mathrm{high}}$ has mean score $0.3$ with range $[0.1626, 1]$. 

The VQE objective is $E_{\eta}(\boldsymbol{\theta}) := \langle 0^{\otimes 8}| U_{\eta}(\boldsymbol{\theta})\dagg H U_{\eta}(\boldsymbol{\theta}) |0^{\otimes 8}\rangle$ for $\eta \in \{ \mathrm{low}, \mathrm{high} \}$.
The exact ground-state energy of (\ref{exp: Hamiltonian}) is $E_0 \approx -13.92$. We run $6$ independent trials for each ans\"atz with the same engineering setting. The final energies are shown in \textbf{Fig.}~\ref{fig: VQE exp}(b). Across all trials, $U_{\mathrm{low}}$ reaches lower energy than $U_{\mathrm{high}}$ as theoretically anticipated. The mean final energies are $\overline{E}_{U_{\mathrm{low}}} = -8.06 \pm 1.2 \, < \, \overline{E}_{U_{\mathrm{high}}} = -1.84 \pm 0.76$. These results support our theoretical studies developed above. \textbf{Code:} \url{https://github.com/hsinyilin19/Observable_Geometry_exp}.

\section{Conclusions}

This study establishes a geometric framework for identifying effective degrees of freedom in VQCs. Theorem~\ref{Thm: disjoint Hermitian orbits} shows that the circuit search space should not be the full group $\U(V)$, but the quotient space $S\backslash \U(V)$, where $S$ removes unitary directions preserving the observable $H$. Theorem~\ref{Thm: Stabilizer decomp} gives a meaningful decomposition of $S$ via the spectral data. 

Via group actions, $S$ is identified as a stabilizer of action $\Ad_{g^{-1}}$. Theorems~\ref{Thm: Lie group homogeneous space},~\ref{Thm: Homogeneous characterization} further identify $S\backslash \U(V)$ with orbit $M_H$, endowing a homogeneous-space structure to $S\backslash \U(V)$. 

We further develop a stabilizer-overlap score measuring how much a circuit generator lies in stabilizer algebra $\mathfrak{s}$, giving a practical criterion for ansatz selection. The VQE experiment confirms that generators with minimal projection onto $\mathfrak{s}$ are more efficient. Thus, for VQAs and quantum architecture search, circuit directions contained in the stabilizer should be removed to avoid wasting optimization resources.

\bibliographystyle{IEEEtran}
\bibliography{references,bib/qml_examples,bib/vqc,bib/explain_qml}

\end{document}